\documentclass[useAMS,usenatbib,onecolumn]{mn2e}
\usepackage{amsmath}
\usepackage{graphicx}
\usepackage{natbib}
\usepackage{color}

\newcommand{\ltsima} {$\; \buildrel < \over \sim \;$}
\newcommand{\gtsima} {$\; \buildrel > \over \sim \;$}
\newcommand{\lta} {\lower.5ex\hbox{\ltsima}}
\newcommand{\gta} {\lower.5ex\hbox{\gtsima}}

\newcommand{\change}[1]{{\textcolor{black}{#1}}}

\def\ln{\mathrm{ln}}

\newcommand{\ba}{\mathbf a}
\newcommand{\data}{\mathbf d}
\newcommand{\x}{\mathbf x}
\newcommand{\y}{\mathbf y}
\newcommand{\z}{\mathbf z}
\newcommand{\bX}{\mathbf X}
\newcommand{\Y}{\mathbf Y}

\newcommand{\calL}{L}
\newcommand{\calP}{{\cal P}}

\newcommand{\A}{{\sf{A}}}
\newcommand{\B}{{\mathbf{B}}}
\newcommand{\C}{{\sf{C}}}
\newcommand{\CXX}{{\sf{C}}_{\rm{XX}}}
\newcommand{\CXY}{{\sf{C}}_{\rm{XY}}}
\newcommand{\CYY}{{\sf{C}}_{\rm{YY}}}
\newcommand{\E}{{\sf{E}}}
\newcommand{\F}{{\sf{F}}}
\newcommand{\G}{{\sf{G}}}
\newcommand{\HH}{{\sf{H}}}
\newcommand{\I}{{\sf{I}}}
\newcommand{\J}{{\sf{J}}}
\newcommand{\K}{{\sf{K}}}
\newcommand{\PP}{{\sf{P}}}
\newcommand{\R}{{\sf{R}}}
\newcommand{\T}{{\sf{T}}}
\newcommand{\U}{{\sf{U}}}
\newcommand{\V}{{\sf{V}}}
\newcommand{\W}{{\sf{W}}}

\title[Generalised Fisher Matrices]{Generalised Fisher Matrices}

\author[A.F. Heavens, M. Seikel, B.D. Nord, M. Aich, Y. Bouffanais,  B.A. Bassett, M.P. Hobson]{A.F. Heavens$^1$\thanks{e-mail:  a.heavens@imperial.ac.uk}, M. Seikel$^2$, B.D. Nord$^3$,  M. Aich$^4$, Y. Bouffanais$^1$, B.A. Bassett$^{5,6,7}$, \\ \\{\LARGE\rm M.P. Hobson$^8$}\\
$^1$  Imperial Centre for Inference and Cosmology, Department of Physics, Imperial College, Blackett Laboratory, \\Prince Consort Road, London SW7 2AZ, U.K.\\
$^2$ The UCT Astrophysics, Cosmology and Gravity Centre, Department of Mathematics and Applied Mathematics,\\
University of Cape Town, Rondebosch 7701, Cape Town, South Africa\\
$^3$ Department of Physics, University of Michigan, Ann Arbor, Michigan, United States\\
$^4$ School of Mathematics, Statistics \& Computer Science, University of KwaZulu-Natal, Durban 4000, South Africa\\
$^5$African Institute for Mathematical Sciences, 6 Melrose Road, Muizenberg, 7945, South Africa\\
$^6$ Department of Mathematics and Applied Mathematics, University of Cape Town, Rondebosch, Cape Town, 7700, South Africa\\
$^7$ South African Astronomical Observatory, Observatory Road, Observatory, Cape Town, 7935, South Africa\\
$^8$ Battcock Centre for Experimental Astrophysics, University of Cambridge, Madingley Road, Cambridge, CB3 0HA}
\date{Accepted ;  Received ; in original form }

\begin{document}
\maketitle

\begin{abstract}
The Fisher Information Matrix formalism \citep{Fisher} is extended to cases where the data is divided into two parts ($\bX,\Y$), where the expectation value of $\Y$ depends on $\bX$ according to some theoretical model, and $\bX$ and $\Y$ both have errors with arbitrary covariance.  In the simplest case, ($\bX,\Y$) represent data pairs of abscissa and ordinate, in which case the analysis deals with the case of data pairs with errors in both coordinates, but $\bX$ can be {\em any} measured quantities on which $\Y$ depends.  The analysis applies for arbitrary covariance, provided all errors are gaussian, and provided the errors in $\bX$ are small, both in comparison with the scale over which the expected signal $\Y$ changes, and with the width of the prior distribution.  This generalises the Fisher Matrix approach, which normally only considers errors in the `ordinate' $\Y$. In this work, we include errors in $\bX$ by marginalising over latent variables, effectively employing a Bayesian hierarchical model,  and deriving the Fisher Matrix for this more general case. The methods here also extend to likelihood surfaces which are not gaussian in the parameter space, and so techniques such as DALI (Derivative Approximation for Likelihoods) can be generalised straightforwardly to include arbitrary gaussian data error covariances. For simple mock data and theoretical models, we compare to Markov Chain Monte Carlo experiments,  illustrating the method with cosmological supernova data.   We also include the new method in the Fisher4Cast software.
\end{abstract}


\section{Introduction}

The Fisher Information Matrix or simply Fisher Matrix has become one of the most widely used statistical tools for forecasting the errors in parameter estimation problems.  It provides lower limits on the variances \change{(through the Cram\' er-Rao inequality)}, and the expected covariances of estimates of model parameters from maximum likelihood, or maximum posterior, techniques, for a given experimental design.  \change{If we further assume gaussianity in two respects: that the data are jointly gaussian-distributed, and that the posterior for the parameters is gaussian, then the Fisher matrix determines the full expected posterior}.  For data pairs $\{X_i,Y_i\}$ with no errors in $X$, the problem was solved many years ago \citep{Fisher}. The main value of the Fisher matrix technique is in being able to obtain error forecasts without any data, real or simulated, and is generally much faster than computing full posterior distributions with simulations \citep{2012ApJ...749...72A, Bassett:2009wr}. It is however only a first step, as it assumes the posteriors are well described by multivariate gaussian distributions, and this may not hold \citep[e.g.,][]{Wolz}, when more sophisticated analysis may be required, but it is still a very valuable tool for experimental design.  Furthermore, more sophisticated forecasts for likelihood surfaces which are non-gaussian in the parameter space now exist \citep{Sellentin}.

From the initial derivations of the Fisher Matrix in the cosmological context \citep{VS96,Tegmark:1997}, we have arrived today at very mature
applications and implementations \citep[e.g.,][]{Bassett:2009wr,   Coe:2009ti, ARefregier:2011ho}.  The Fisher Matrix has been useful in proposals and projections for surveys, such as for the Cosmic Microwave Background \citep{1997astro.ph..7265T}, spectroscopic galaxy surveys \citep{BigBoss}, the Dark Energy Survey \citep{2005astro.ph.10346T}, large-scale structure \citep{2009PhRvD..79f3009C}, and in the broader discussion of the investigation of Dark Energy \citep{2006astro.ph..9591A} and estimation of neutrino masses with the future European Space Agency Euclid mission \citep{Kitching2008}.

For the purposes of review and later reference in this work, we
summarise the basic Fisher Matrix formalism.
We begin with the likelihood of a set of data, ${\mathbf d}$ given (or conditional
upon) a set of model parameters, represented by a vector $\btheta$: 
$p({\mathbf d}|{\btheta})$.  In the simplest case, $\mathbf d$ represents only the ordinates, $\Y$.  Later in the paper, we will take it to be the union of the ordinates and any other measured quantities on which $\Y$ depends, such as abscissa values, and which may be subject to error.  In practice what is typically required is the posterior distribution of $\btheta$, given the data $\data$. Assuming an uninformative prior on the parameters, $p(\btheta)=$ constant, Bayes' Theorem implies $p({\btheta}|{\mathbf d})\propto p({\mathbf d}|{\btheta})=L$, the likelihood.  The
log-likelihood is then Taylor-expanded about its maximum.  The
first term is a constant, irrelevant for the discussion of parameter
constraint forecasts; the second term is the first derivative, which
vanishes at the point of maximum likelihood; the third term is the
Hessian (curvature matrix) of the likelihood, and is the term whose
ensemble average (over the data) gives the Fisher Matrix:
\begin{equation}
F_{\alpha\beta} = -\left\langle \frac{\partial^2 \ln L}{\partial
    \theta_\alpha \partial \theta_\beta} \right\rangle,
\end{equation}
where $\alpha$ and $\beta$ label the parameters.
For the case of a gaussian likelihood, this is analytically computable, and can depend only on the expectation values of the data, $\mu(\btheta)\equiv \langle {\bf d}(\btheta)\rangle$, and the covariance, $\C(\btheta)\equiv\langle ({\bf d}-\bmu)^T({\bf d}-\bmu)\rangle$.  This results in the following form for the Fisher Matrix \citep{Tegmark:1997}.
\begin{equation}
\F_{\alpha\beta} =
\frac{1}{2}{\rm{Tr}}\left[\C^{-1}\C_{,\alpha}\C^{-1}\C_{,\beta} +
  \C^{-1}(\bmu_{,\alpha}\bmu_{,\beta}^T+\bmu_{,\beta}\bmu_{,\alpha}^T)\right]. \label{eqn:originalFM}
\end{equation}
An early example of dealing with errors in both variables was straight-line fitting,
where both the statistics and astronomy communities used either {\em ad hoc} choices for the axis, or ultimately
arbitrary combinations e.g., the bisector or the average of the
one-dimensional fits on either axis.  The evolution to two-dimensional or joint-distribution fitting was
accompanied by a slow transition to the Bayesian perspective
\citep{Gull:1989uy}.  New tools for fitting
data in the presence of two-dimensional errors have been developed and
used to extract improved cosmological constraints from supernovae populations
\citep{2011MNRAS.418.2308M}. Here, we develop the application of
two-dimensional errors in the predictive Fisher Matrix formalism itself, \change{but the formalism can treat more general cases where the signal depends on arbitrary extra parameters} . 
For pedagogical discussions of straight-line fitting and Bayesian
approaches to fitting, see for example
\cite{Hogg, DAgostini:2005we,Kelly}.

The remainder of the paper is organized as follows:
\S\ref{sec:formalism} describes the formal derivation of the generalized Fisher matrix
for the case of dependence of $\Y$ on an arbitrary set of gaussian-distributed variables $\bX$;  \S \ref{sec:examples}  describes an application of this formalism to a
particular experiment, with tests on simulated data. We present conclusions in \S\ref{sec:conclusion}.  \change{For the reader who is interested only in the application of the result, this is effected by simply replacing the covariance matrix $\C$ in equation (\ref{eqn:originalFM}) by the matrix $\R$ computed in equation (\ref{Rmatrix})}.

\section{Formalism of the Extension}\label{sec:formalism}
Throughout this paper, we follow the formalism and notation of \cite{Bassett:2009wr}.
In this method, we use a Taylor expansion of the log-likelihood, and
derive the generalised Fisher Matrix from first principles.  The general aim is to find an expression for the Fisher Matrix for an
experiment with gaussian errors in $\bX$ and $\Y$,  arbitrary
correlations of errors (i.e. errors in $Y_i$ can be correlated with
errors in $X_j$, for any $i,j$).  \change{As previously mentioned, the formalism covers the case when $\bX$ represents the abscissa values of the data points, but it need not, and the extra variables may not be associated with an individual $Y_i$ at all.}

\subsection{General Method with $X$-$Y$ Covariance}

Let the set of measurements be $\{X_i\},\{Y_j\}$, with $i=1,\ldots M$ and $j=1,\ldots N$.  In the simplest case, $M=N$ and the dataset is a set of ($X,Y$) data pairs, but this is not necessary; all that is required is that there a model which returns the expectation value of $\Y$ as a function of $\bX$, and which in general will depend also on some model parameters, represented collectively by $\btheta$, being a vector $\theta_{\alpha}$ with $\alpha=1, \ldots P$.  It is the posterior probability of $\btheta$ which we wish to calculate. We give an example later.

We assume $\bX$ and $\Y$ have Gaussian errors, around true values $\x$, $\y$, with a covariance matrix $\C$.   $\x$ and $\y$ are
not directly observed.  This amounts to a hierarchical model, where the observables $\bX, \Y$ depend on some unobservable latent variables $\x$, which are essentially nuisance parameters.  The $\y$ are not independent nuisance parameters as they are assumed to be related precisely by a theoretical model $\y=\bmu(\x)$, which also depends on $\btheta$.
We seek the posterior $p(\btheta|\bX,\Y)$.  With a uniform prior for $\btheta$, this is proportional to the likelihood $L = p(\bX,\Y|\btheta)$. 
We write this as the marginalised distribution over $\x$ and $\y$ as
\begin{eqnarray}
L &=& \int p(\bX,\Y,\x,\y|\btheta) \, d\x\, d\y 
= \int p(\bX,\Y|\x,\y,\btheta)p(\x,\y|\btheta)  \, d\x\, d\y\\\nonumber
&=&
\int p(\bX,\Y|\x,\y,\btheta)p(\y|\x,\btheta)p(\x|\btheta)  \, d\x\, d\y 
\end{eqnarray}
\noindent where we have expanded the condition to include the latent
variables, and then further expanded the condition of $p(\y)$ to
include $\x$.

We integrate over $\y$ using a delta function,
$p(\y|\x,\btheta)=\delta(\y-\bmu(\x))$, and assume for now a uniform prior for $\x$:
\begin{equation}
L = \int   p(\bX,\Y|\x,\bmu(\x),\btheta)  \, d^M\x.
\end{equation}
At the cost of some algebraic complexity, we can introduce an informative prior (parent distribution) for $\x$.  In Appendix B, we generalise the analysis by assuming a gaussian population prior $p(\x)$, and show that we recover the simpler result obtained in the main text in the limit that the errors in $\x$ are small enough that the prior can be considered constant across the error range of individual data points. \change{Note that formally we assume the prior is independent of the model parameters, but in the limit discussed in the main text, any such dependence does not affect the result.} See \cite{Gull:1989uy} and \cite{Kelly} for further discussion of these points. In this paper we are not explicitly concerned with biases, but it is important to note Gull's point that estimates of parameters, such as the slope of a straight-line fit with errors in both coordinates, will be biased, even with an informative prior, unless the width of the prior is a hyperparameter that is marginalized over.  No doubt similar considerations will be important in applications of the more complicated situation considered here.

Next, we make the critical assumption that we can truncate at the
linear term of the Taylor expansion of $\bmu$:
\begin{equation}
\bmu(\x)=\bmu(\bX)+\T(\bX)\,(\x-\bX),
\end{equation}
where
\change{
\begin{equation}
\T_{ij} \equiv \left.\frac{\partial \mu_i}{\partial x_j}\right |_{\x=\bX}.
\end{equation}
}
\change{In the case when $\bX$ represents the abscissa values, we would expect $\T$ to be diagonal.}

We are essentially assuming that the function $\bmu(\x)$ is linear
across the width of the gaussian error distribution of $\x$, and this allows the likelihood to be integrated analytically, as it is simply a gaussian integral:
\begin{equation}
L \propto \int
\frac{1}{\sqrt{\det\C}}\exp\left(-\frac{Q}{2}\right)\,d\x \label{eqn:likelihood_gaussian}
\end{equation}
where $Q \equiv ({\bf Z}-\z)^T\C^{-1}({\bf Z}-\z)$, and $\z$ and ${\bf Z}$ are $M+N$-dimensional vectors:  $z_i=x_i$ and $Z_i=X_i$  for $i\le M$, $Z_{M+j}=Y_j$ and $z_{M+j}=\mu_j(\bX)+[\T(\bX)(\x-\bX)]_j$.

The covariance matrix of the data can be written in block form as
\begin{equation}
\C = \bordermatrix{~ & X & Y \cr
                  X & \CXX & \CXY \cr
                  Y & \CXY^T & \CYY \cr}.
\end{equation}
Note that $\C_{\rm{XY}}$ is not symmetrical, nor invertible or even square in general; although $\C_{\rm{XX}}$ and $\C_{\rm{YY}}$ are.    The covariance matrix may include a number of elements, such as intrinsic scatter and measurement noise, with individual covariance matrices adding to give the final $\C$. The inverse of $\C$ is
\begin{equation}
\C^{-1} = \left(\begin{matrix}
\G & -\HH \cr
-\HH^T & \E
\end{matrix} \right)
\end{equation}
where
\begin{eqnarray}
\G &=& \C_{\rm{XX}}^{-1}+ \C_{\rm{XX}}^{-1}\C_{\rm{XY}}\E \C_{\rm{XY}}^T \C_{\rm{XX}}^{-1}\cr
\HH &=& \C_{\rm{XX}}^{-1}\C_{\rm{XY}}\E\cr
\E &=& (\C_{\rm{YY}}-\C_{\rm{XY}}^T\C_{\rm{XX}}^{-1}\C_{\rm{XY}})^{-1}.
\end{eqnarray}
Defining \change{$\tilde \x \equiv \bX-\x$}, and $\tilde \Y\equiv \Y-\bmu(\bX)$, we collect together the terms as follows:
\change{
\begin{equation}
Q = \tilde \x^T \G \tilde \x +(\tilde \Y+\T\tilde \x)^T \E (\tilde \Y+\T\tilde \x)-
\tilde \x^T \HH (\tilde \Y+\T\tilde \x)- (\tilde \Y+\T\tilde \x)^T \HH^T \tilde \x.
\end{equation}
}
$Q$ has the  quadratic form
\begin{equation}
Q = \tilde \x^T \A \tilde \x - \B^T\tilde \x  - \tilde \x^T\B+ Q', \label{eqn:qdefinition}
\end{equation}
where
\change{
\begin{eqnarray}
\A &=& \G + \T^T\E\T - \HH\T - \T^T\HH^T\nonumber\\
\B &=& (\HH-\T^T\E)\tilde \Y \equiv \PP\tilde \Y\nonumber\\
Q' &=&  \tilde \Y^T\E \tilde \Y.
\end{eqnarray}
}
With the definition of $Q$ in Eqn.~\ref{eqn:qdefinition}, the gaussian integral of Eqn.~\ref{eqn:likelihood_gaussian} can be performed, using
\begin{equation}
\int \exp\left({-\frac{1}{2}\tilde \x^T\A\tilde \x +\B^T\tilde
  \x}\right) d\tilde \x =
\frac{(2\pi)^{N/2}}{\sqrt{\det{A}}}\exp\left({\frac{1}{2}\B^T\A^{-1}\B}\right),
\end{equation}
\noindent and noting that $Q'$ is independent of $\tilde x$.
The likelihood then simplifies after a few lines of algebra to
\begin{equation}
L \propto \frac{1}{\sqrt{\det\A\det\C}}\exp\left(-\frac{1}{2} \tilde \Y^T \R^{-1} \tilde \Y\right),
\end{equation}
where the inverse of the marginal covariance matrix of $\tilde \Y$ is 
\change{\begin{equation}
\R^{-1} = \E-\PP^T\A^{-1}\PP.
\end{equation}
We use the Woodbury formula \citep{1950woodbury}
\begin{equation}
(\K + \U\W\V)^{-1} = \K^{-1}- \K^{-1}\U(\W^{-1}+\V\K^{-1}\U)^{-1}V\K^{-1}
\end{equation}
to obtain after some algebra}
\change{
\begin{equation}
\R = \C_{\rm{YY}}-\C_{\rm{XY}}^T\T^T -\T\C_{\rm{XY}}+\T\C_{\rm{XX}}\T^T,
\label{Rmatrix}
\end{equation}
}
which is the key result of the calculation.  We can also simplify the
pre-factor, $\det\A\det\C = \det\R$  (see Appendix~\ref{app:car} for
the proof).  Thus
\begin{equation}
L \propto \frac{1}{\sqrt{\det\R}}\exp\left(-\frac{1}{2}\tilde \Y^T \R^{-1} \tilde \Y\right).
\label{LikelihoodR}
\end{equation}
We see that this looks just like a normal gaussian (in terms of data) likelihood, but with the covariance matrix $\C$ ($\CYY$ in our current notation) replaced by $\R$.  Hence to compute the Fisher matrix, we can use the standard formula found in Eqn.~\ref{eqn:originalFM} and Eqn. 15 of \cite{Tegmark:1997}, and simply replace $\C$ by $\R$:
\begin{equation}\label{Fnew}
\F_{\alpha\beta} = \frac{1}{2}{\rm{Tr}}\left[\R^{-1}\R_{,\alpha}\R^{-1}\R_{,\beta} + \R^{-1}(\bmu_{,\alpha}\bmu_{,\beta}^T+\bmu_{,\beta}\bmu_{,\alpha}^T)\right].
\end{equation}
This is the main result of this paper.
Note that $\R$ depends not only on the standard covariance, but also on the covariance in the
independent variable, $\C_{\rm{XX}}$, the meta-covariance, $\C_{\rm{XY}}$, and
the first partial derivatives of the model function $\bmu$.  In the case of uncorrelated data pairs, the result reduces to that found in March et al (2011).  For the simple case of no correlations between $\bX$ and $\Y$ values $\R=\CYY+\T^T\CXX\T$, and with diagonal covariance matrices $\C_{\rm{YY}}$ and  $\C_{\rm{XX}}$ we recover the propagation of error result that the variance of $f\equiv Y-\mu(X)$ for each data point is effectively
\begin{equation}
\sigma_f^2 = \sigma_{\rm Y}^2 + \mu'(X)^2\,\sigma_{\rm X}^2,
\end{equation}
where $\mu' = \partial\mu/\partial x$ and $\C$ can be replaced in the standard Fisher expression (\ref{eqn:originalFM}) by a diagonal $ N \times N$ matrix with these enhanced entries. 

We now briefly make a few key observations.  First, when the
derivatives of the model function are zero ($\T=0$), then the latent variable
$\x$ has no bearing on $\R$, and we recover the usual formula for the
Fisher Matrix: when $\T=0$, $\R = \C_{\rm{YY}}$.  Also, in the limit of
infinitesimal errors in $\bX$, we recover the usual Fisher matrix
formula.  As remarked earlier, if the errors in $\bX$ and $\Y$ are uncorrelated, and in the limit that the errors in $\bX$ are small in comparison with the width of the prior, we recover the result obtained from propagation of errors, namely that the variance of $\Y$ is effectively increased 
from $\sigma_{\rm Y}^2$ to $\sigma_{\rm Y}^2+\mu'^2\sigma_{\rm X}^2$.  Also, although the main focus of the paper has been on the Fisher matrix, the expression for the likelihood itself (equation \ref{LikelihoodR}) can be used without the usual interpretation that it is gaussian in the parameter space, to make predictions for the shape of the likelihood surfaces beyond ellipses.  Thus the technology of DALI \citep{Sellentin} can be generalized straightforwardly by replacing the data covariance matrix by $\R$.  Finally, even if the covariance matrix of the data (the
original $\C$, which is $\C_{\rm{YY}}$) is independent of the parameters, $\R$ is not, because in general  $\T$ does depend on the parameters.

\section{Example Application}\label{sec:examples}

As an example for illustration, consider the Type 1A supernova Hubble diagram, which consists of data pairs corresponding to the redshift of the host galaxy of each supernova, and its apparent brightness.  In the case presented here $\bX$ and $\Y$ have the same length, and represent the redshifts and distance moduli of the supernovae. Various corrections, based on colour and the timescale of decline of the light curve (`stretch'), are applied such that these supernovae act as standard candles with a small dispersion of around 10\%.  Colours and stretch could be added to $\bX$, in which case $M\ge 3N$, but $\bX$ could also include variables which are not associated with a single $Y_j$ value (e.g. instrumental calibration).   The $\Lambda$ cold dark matter model plus empirical corrections for colour and stretch then relate $\Y$ to $\bX$, dependent on parameters of interest, such as the matter and dark energy content. See \cite{Mandel} for a full Bayesian hierarchical model description, and \cite{2011MNRAS.418.2308M} for a principled analysis of data, and further discussion of background. Redshift errors obtained from spectroscopy are negligibly small, but if they are photometric redshifts, based on broad-band colours of the host galaxy, then two complications arise.  One is that the redshift errors may be large (typically around 5 or 10\% for 5-band photometry).  The second is that errors in the photometry (such as zero-point errors) will introduce errors in the redshifts, but could also affect the colour corrections for the supernovae themselves.  This potentially couples the errors in $X$ and $Y$ for a given data pair.  \change{In the rare case of a galaxy with multiple supernovae, a mis-estimation of host galaxy extinction would couple redshift errors, as well as apparent brightness, of the affected supernovae.}  \cite{Kim} investigated correlations between redshift and magnitude errors in photometric surveys, and found rather variable correlation coefficients between about 0.35 and 0.95. 
\begin{figure}
\begin{center}
\includegraphics[width = 12cm]{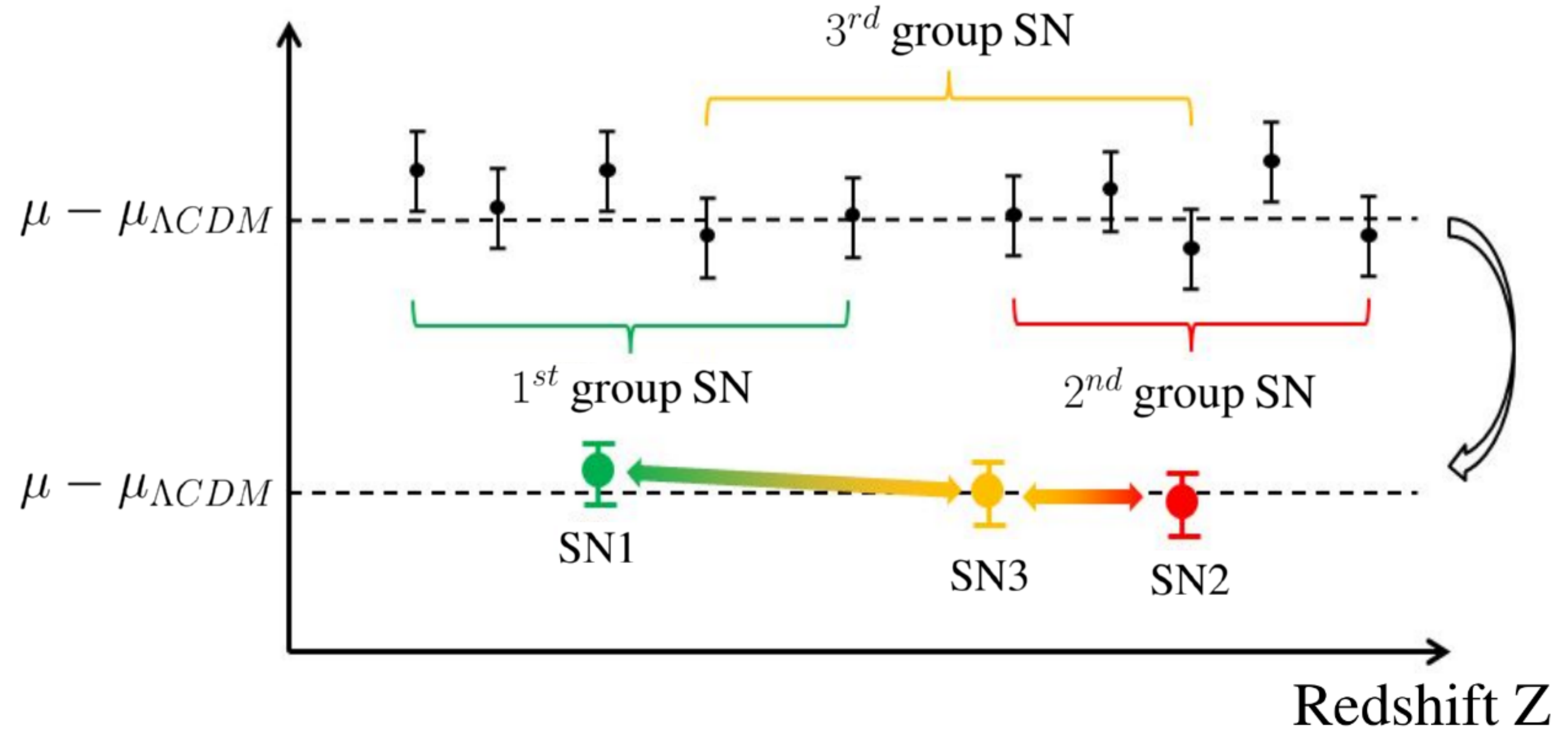}
\end{center}
\caption{A scenario in which the formation of overlapping weighted combinations of the original data may lead to correlations between $X$ and $Y$ values of different pairs.  Here, the $Y$ values have been adjusted to the theoretical curve for a fiducial set of model parameters, which is a function of $X$, so errors in $X$ propagate into $Y$, and the weighting then mixes different $Y$ values.  This then correlates both $X$ and $Y$ values from different pairs. $\mu$ and $\mu_{\Lambda CDM}$ are the measured and theoretical distance moduli, with the theoretical model chosen for illustration to be the $\Lambda CDM$ concordance model.}
\label{SNsamples}
\end{figure}

A scenario which could couple the errors in $X$ and $Y$ for different data pairs arises if one takes weighted averages of the data.  This one might do in order to make the errors closer to gaussian, as we do not know the error distribution for individual supernovae.  If this is done with overlapping sub-samples, to maintain a good sampling in redshift (see Fig. \ref{SNsamples}), then the errors will be coupled.  Furthermore, if the $Y$ values are referred to a fiducial model (such as the standard cosmological model), as shown, then this involves dividing by a function of the supernova redshift, which then couples the errors in $X$ to the errors in $Y$ across different (weighted) data pairs.   So we see in this example how one can get full covariance between $X$ and $Y$ sets, with non-zero off-diagonal terms of all types.  

\begin{figure}
\begin{center}
\includegraphics[width = 12cm]{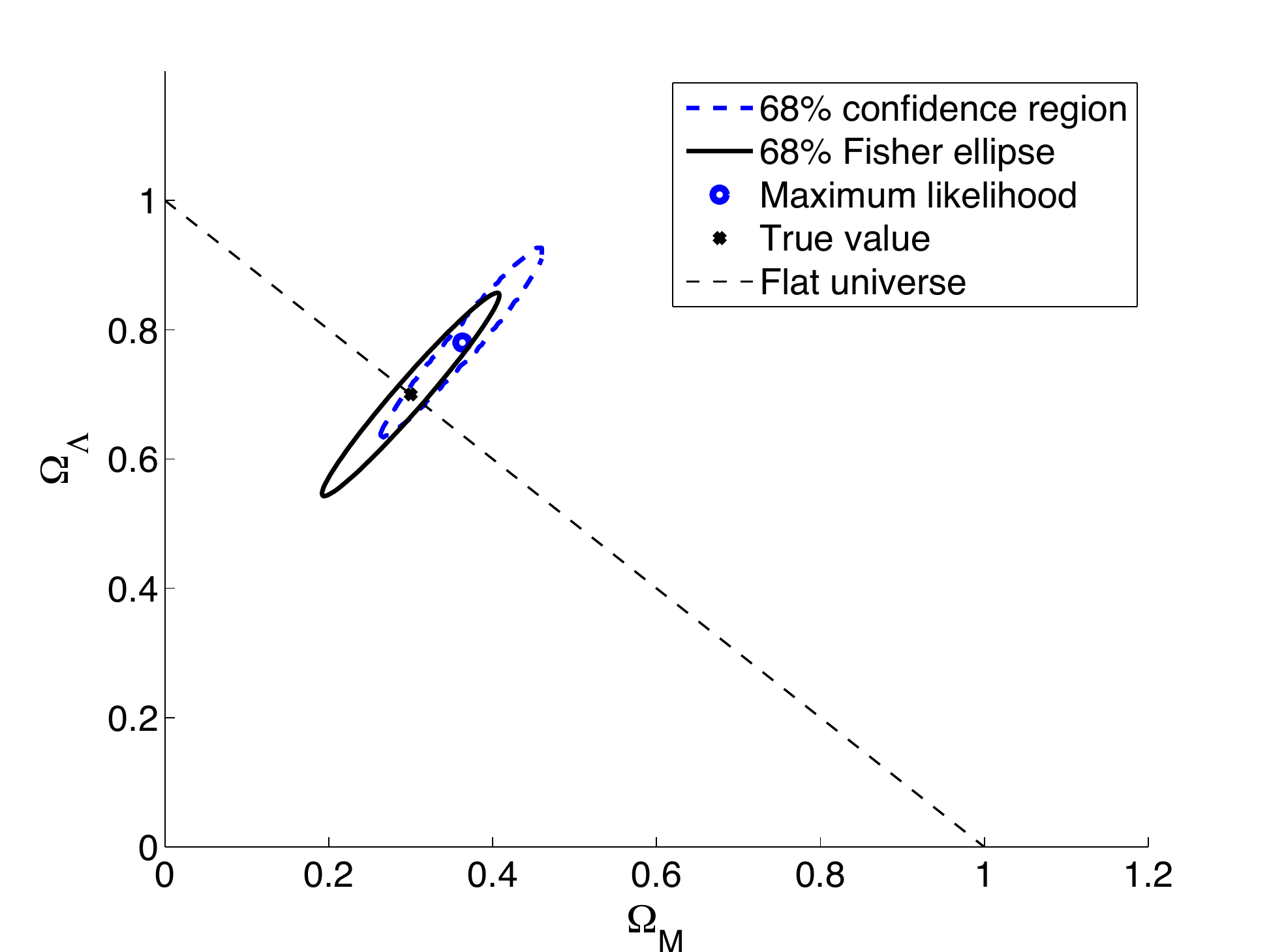}
\end{center}
\caption{Generalised Fisher Matrix calculations compared with MCMC results from simulated supernova data generated with correlations between $\bX$ and $\Y$ values in each data pair.  The likelihood is accurately a bivariate gaussian for this example, and there is good agreement in the shape, size and orientation of the ellipses, with the actual likelihood offset from the true solution in accordance with expectation.}
\label{SNsim}
\end{figure}

To illustrate results using the generalised Fisher matrix, we have simulated supernovae with correlated errors in redshift and distance modulus,  obtaining an estimate of the posterior for the matter density parameter and cosmological constant, using Markov Chain Monte Carlo techniques.  For illustration we show the simplest non-trivial case, where 200 supernovae are drawn from a uniform distribution of redshifts $z$ in the range $0<z<1.1$, each having uncorrelated gaussian errors of 0.1 in distance modulus and 0.01 in $z$; more complicated examples look essentially the same.   Fig. \ref{SNsim}  shows the comparison of the MCMC error ellipse with the expected error contours from the generalised Fisher Matrix technique, showing good agreement.


\section{Conclusions}\label{sec:conclusion}

In this paper we have considered the Fisher Information Matrix where some subset of the data ($\Y$) depends via a theoretical model $\langle \Y\rangle = \bmu(\bX,\btheta)$ on some other set of measured variables ($\bX$), and a set of model parameters $\btheta$ whose posterior distribution is desired.  $\bX$ and $\Y$ are assumed to have gaussian errors which can have arbitrary covariance.  This includes as a subset the case of ($X,Y$) data pairs with errors in both coordinates,  with correlations between one independent variable and a different dependent variable, but the analysis is more general, and $\bX$ can included any other measured quantities. The main result, equation (\ref{Fnew}), is similar to the standard Fisher matrix, but with the covariance matrix replaced by a more complicated matrix (\ref{Rmatrix}) derived from the expanded covariance matrix of all variables, and the partial derivatives of the expected signals with respect to the dependent variables.  The result is valid for situations where two conditions hold: the first is that a Taylor expansion of the expected signal to linear order is valid across the gaussian error of the independent variables; the second is that the errors in the independent variables are small compared with the width of the prior distribution.  At the price of some complexity, we present a perturbative correction when the latter condition does not hold.  In the case when the errors are uncorrelated between data pairs, the result reduces to the result one obtains from propagation of errors, where the variance of the dependent variable is increased from $\sigma_{\rm Y}^2$ to $\sigma_{\rm Y}^2+(\partial \mu/\partial x)^2\sigma_{\rm X}^2$.  Since we compute the likelihood itself, it may be used to evaluate the expected likelihood surface when it is not gaussian in the parameter space, straightforwardly generalizing the DALI technique of \cite{Sellentin}.  Finally, the generalised Fisher Matrix will be implemented in the Fisher4Cast software, available at http://www.mathworks.com/matlabcentral/fileexchange/20008-fisher-matrix-toolbox-fisher4cast.

\noindent{\bf Acknowledgments}\\
We are grateful to the organisers of the Cape Town International Cosmology School, where this work started as a student project, to Roberto Trotta, Daniel Mortlock and Andrew Jaffe for useful discussions, and to the anonymous referee for very helpful comments and suggestions.

\appendix
\section{Proof that $\det\C \det\A = \det\R$}\label{app:car}
\change{
With $\det\C = \det(\C_{\rm{XX}})\det(\C_{\rm{YY}}-\C_{\rm{XY}}^T\C_{\rm{XX}}^{-1}\C_{\rm{XY}})=\det(\C_{\rm{XX}})\det(\E^{-1})$, we have, reversing the order of the determinants,
\begin{equation}
\det\C\det\A=\det(\E^{-1})\det(\C_{\rm{XX}}) 
\det\left[\C_{\rm{XX}}^{-1} + (\C_{\rm{XX}}^{-1}\C_{\rm{XY}}-\T^T)\E(\C_{\rm{XX}}^{-1}\C_{\rm{XY}}-\T^T)^T\right]
\end{equation}
Now, since $\det\U\det\V=\det(\U\V)$ for any square matrices,
\begin{equation}
\det\C\det\A=\det(\E^{-1})\det\left[{\sf{I}} + (\C_{\rm{XY}}-\C_{\rm{XX}}\T^T)\E(\C_{\rm{XX}}^{-1}\C_{\rm{XY}}-{\T}^T)^{\it T}\right].
\end{equation}
Now we use Sylvester's Determinant Theorem, $\det(\sf{I}+\U\V)=\det(\sf{I}+\V\U)$ where we take $\V = \E(\C_{\rm{XX}}^{-1}\C_{\rm{XY}}-\T^T)^T$:
\begin{equation}
\det\C\det\A=\det(\E^{-1})\det\left[{\sf{I}} + \E(\C_{\rm{XX}}^{-1}\C_{\rm{XY}}-\T^T)^{\it T}(\C_{\rm{XY}}-\C_{\rm{XX}}\T^T)\right].
\end{equation}
Using  $\det\U\det\V=\det(\U\V)$ again, and expanding $\E^{-1}$,
\begin{eqnarray}
\det\C\det\A&=&\det\left[\E^{-1}+ (\C_{\rm{XX}}^{-1}\C_{\rm{XY}}-\T^T)^T(\C_{\rm{XY}}-\C_{\rm{XX}}\T^T)\right]\nonumber\\
&=& \det\left[\C_{\rm{YY}}-\C_{\rm{XY}}^T\C_{\rm{XX}}^{-1}\C_{\rm{XY}} + (\C_{\rm{XY}}^T\C_{\rm{XX}}^{-1}-\T)(\C_{\rm{XY}}-\C_{\rm{XX}}\T^T)\right]\nonumber\\
&=& \det\left[\C_{\rm{YY}}- \C_{\rm{XY}}^T \T^T -\T\C_{\rm{XY}}+\T\C_{\rm{XX}}\T^T\right]=\det\R.
\end{eqnarray}
}

\section{Generalisation to non-uniform prior, or parent distribution}

We now generalise the method to apply to cases where the prior in $\x$ is not uniform.  We illustrate this with a simplifying assumption that the prior is a gaussian of specified width, and demonstrate that in the limit of a prior width which is much larger than the errors in $\x$, we recover the results in the main text, and we expect this to hold for any broad prior.   We can consider a prior which is dependent on each point, with a mean vector $\ba$ and variance $\Sigma$ (we assume that $\Sigma$ is a diagonal matrix).  In the normal case where the abscissa values are drawn from the same distribution, then all elements of $\ba$ are identical, and $\Sigma$ is proportional to the identity matrix.

Assuming a gaussian prior
\begin{equation}
p(\x) \propto \exp\left[-\frac{1}{2}(\x-\ba)^T \Sigma^{-1} (\x-\ba) \right]
\end{equation}
we get for the posterior
\begin{equation}
\calP \propto \int \frac{1}{\sqrt{\det\C}}\exp\left\{-\frac{1}{2}
\left[ Q + (\x-\ba)^T \Sigma^{-1} (\x-\ba) \right]\right\}\,d^N\x \,,
\end{equation}
Defining $\tilde\bX\equiv \bX-{\bf a}$, we get
\begin{equation}
Q + (\x-\ba)^T \Sigma^{-1} (\x-\ba) 
=
\tilde\x^T \tilde\A \tilde\x - 2\tilde\B^T \tilde\x + 
\tilde\bX^T\Sigma^{-1}\tilde\bX + Q'
\end{equation}
where
\begin{eqnarray}
\tilde\A &=& \A + \Sigma^{-1}\\
\tilde\B^T &=& \B^T \,+\tilde\bX^T \Sigma^{-1}.
\end{eqnarray}
We perform the gaussian integral as before, finding 
\begin{equation}
\label{likelihood}
\calP \propto \frac{1}{\sqrt{\det\C \det\tilde\A}} 
  \exp\left[-\frac{1}{2} \left(-\tilde\B^T \tilde\A^{-1} \tilde\B +
  \tilde\bX^T \Sigma^{-1} \tilde\bX  + Q'\right)\right].
\end{equation}
In the case when the prior in $\x$ is informative, then there is information in the values of $\bX$, so the data vector should include both $\bX$ and $\Y$.  The likelihood is then
\begin{equation}
\calP \propto \frac{1}{\sqrt{\det\C \det\tilde\A}} 
  \exp\left(-\frac{1}{2}Q_{\rm YX}\right)
\end{equation}
where
\begin{equation}\label{QYX}
Q_{\rm YX} = (\tilde\Y,\tilde\bX)^T \J (\tilde\Y,\tilde\bX)
\end{equation}
and, collecting terms and using the Woodbury identity again, we find
\begin{equation}\label{CovYX}
\J = \left(\begin{matrix}
\E-(\HH^T-\E\T)(\A+\Sigma^{-1})^{-1}(\HH-\T^T\E) & (\HH^T-\E\T)(\A+\Sigma^{-1})^{-1}\Sigma^{-1} \cr 
\Sigma^{-1}(\A+\Sigma^{-1})^{-1}(\HH-\T^T\E) & (\Sigma+\A^{-1})^{-1}
\end{matrix} \right).
\end{equation}
In the limit of an infinitely broad prior, we see that, as expected, $\tilde\bX$ contains no useful information, and the likelihood depends only on $\tilde\Y$, with the quadratic simplifying to
$Q_{\rm YX} \rightarrow Q_{\rm Y} \equiv \tilde\Y^T \R \tilde\Y$, and
as expected, we recover the results of the main text.

To investigate departures from the main text result, we consider terms linear in
$\Sigma^{-1}\A^{-1}$. This approximation only makes sense if
\begin{equation}\label{approxcondition}
\lim_{n\to\infty}\left(\Sigma^{-1}\A^{-1}\right)^n = 0 \,.
\end{equation}
As $\Sigma$ is a diagonal matrix, the elements of the matrix
$(\Sigma^{-1}\A^{-1})^n$ are given by
\begin{eqnarray}
\left[\left(\Sigma^{-1}\A^{-1}\right)^n\right]_{ij} &=& 
\left(\left[\Sigma^{-1}\right]_{ii}\left[\A^{-1}\right]_{ii}\right)^{n-1}
\left[\Sigma^{-1}\right]_{ii}\left[\A^{-1}\right]_{ij} \nonumber\\
&=& \left(\left[\A^{-1}\right]_{ii}/\Sigma_{ii}\right)^{n-1}
\left[\A^{-1}\right]_{ij}/\Sigma_{ii}
\end{eqnarray}
Thus condition \eqref{approxcondition} is fulfilled if
\begin{equation}
\left[\A^{-1}\right]_{ii} \ll \Sigma_{ii}
\end{equation}
for all $i$.
We will assume this and neglect higher order terms in
$\Sigma^{-1}\A^{-1}$. Then we can approximate $\tilde\A^{-1}$ by
\begin{eqnarray}
\tilde\A^{-1} &=& \left(\A + \Sigma^{-1}\right)^{-1} \\\nonumber
&=& \A^{-1} \left(\I + \Sigma^{-1}\A^{-1}\right)^{-1}
\simeq \A^{-1} \left(\I - \Sigma^{-1}\A^{-1}\right)
\end{eqnarray}
Inserting this result in equation \eqref{likelihood}, we get
\begin{equation}
\calP \propto \calL_0 \calL_1
\end{equation}
with
\begin{equation}
\calL_0 =
\frac{1}{\sqrt{\det\C\det\A}}\exp\left(-\frac{1}{2}\tilde \Y^T
\R^{-1} \tilde \Y\right)
\end{equation}
and
\begin{equation}
\calL_1 =
\frac{1}{\sqrt{\det\left(\I + \Sigma^{-1}\A^{-1}\right)}} 
\exp\left[-\frac{1}{2}\left(\A^{-1}\B + \tilde\bX\right)^T 
\Sigma^{-1}\left(\A^{-1}\B + \tilde\bX\right)\right].
\end{equation}
$\calL_0$ is the zeroth order result from the main text.

The Fisher matrix is then given by
\begin{equation}
\F_{\alpha\beta} = \F_{\alpha\beta}^{(0)} +  \F_{\alpha\beta}^{(1)}
\end{equation}
with
\begin{equation}
\F_{\alpha\beta}^{(i)} = -\left\langle
\frac{\partial^2\ln\calL_i}{\partial\theta_\alpha\partial\theta_\beta}
\right\rangle \qquad i=0,1.
\end{equation}
We already know the result for $\F_{\alpha\beta}^{(0)}$, so we just need to calculate the first-order term:
\begin{equation}
\F_{\alpha\beta}^{(1)} = \left\langle
\frac{\partial^2}{\partial\theta_\alpha\partial\theta_\beta}
\left[
\frac{1}{2}\ln\det\left(\I + \Sigma^{-1}\A^{-1}\right) +
\frac{1}{2}\left(\A^{-1}\B + \tilde\bX\right)^T 
\Sigma^{-1}\left(\A^{-1}\B + \tilde\bX\right)\right]
\right\rangle.
\end{equation}
Using the approximation
\begin{equation}
\ln\det\left(\I + \Sigma^{-1}\A^{-1}\right) 
= \rm{Tr}\,\ln\left(\I + \Sigma^{-1}\A^{-1}\right)
\simeq \rm{Tr}\left(\Sigma^{-1}\A^{-1}\right)
\end{equation}
and with $\langle \tilde\Y\rangle = 0$, $ \langle \tilde\Y \tilde\Y^T \rangle = \R$, and $
\tilde\Y_{,\alpha} = -\bmu_{,\alpha}$
we find after some tedious calculations
\begin{eqnarray}
\F_{\alpha\beta}^{(1)} &=&
\frac{1}{2}{\rm{Tr}}\left[ \Sigma^{-1}\{\A^{-1}\}_{,\alpha\beta} +
\left\{\left(\HH^T-\E\T\right) \A^{-1}\Sigma^{-1}\A^{-1} 
\left(\HH-\T^T\E\right)\right\}_{,\alpha\beta} \R \right]
\nonumber\\ &&
 {} -\tilde\bX^T \Sigma^{-1} \left\{\A^{-1} 
\left(\HH-\T^T\E\right) \bmu \right\}_{,\alpha\beta}
+ \tilde\bX^T \Sigma^{-1} 
 \nonumber\\ &&
 \left\{\A^{-1} \left(\HH-\T^T\E\right) \right\}_{,\alpha\beta} \bmu 
\label{Fisher}
{} + \bmu_{,\alpha}^T \left(\HH^T-\E\T\right)
\A^{-1}\Sigma^{-1}\A^{-1} \left(\HH-\T^T\E\right) \bmu_{,\beta}\,.
\end{eqnarray}
As $\{\A^{-1}\}_{,\alpha} = -\A^{-1}\A_{,\alpha}\A^{-1}$, each term in
\eqref{Fisher} contains the factor $\Sigma^{-1}\A^{-1}$, so $\F_{\alpha\beta}^{(1)}$  gives the first-order corrections
in terms of this parameter. 


\bibliographystyle{chicago}

\begin{thebibliography}{}
\bibitem[\protect\citeauthoryear{Acquaviva {et~al.}}{2012}]{2012ApJ...749...72A}
Acquaviva V., Gawiser E., Bickerton S.J., Grogin N.A., Guo Y., Lee S.-K., 2012, The Astrophysical Journal, 749, 72
\bibitem[\protect\citeauthoryear{Albrecht et~al.}{2006}]{2006astro.ph..9591A}
Albrecht A., Bernstein G., Cahn R., Freedman W.L., Hewitt J., Hu W., Huth J., Kamionkowski M., Kolb E.W., Knox L., Mather J.C., Staggs S., Suntzeff N.B., 2006, arXiV:0609591
\bibitem[\protect\citeauthoryear{Bassett {et~al.}}{2009}]{Bassett:2009wr}
Bassett B.A., Fantaye Y., Hlozek R., Kotze J., 2009, arXiv.org, astro-ph.CO
\bibitem[\protect\citeauthoryear{Coe}{2009}]{Coe:2009ti}
Coe D., 2009, Arxiv preprint arXiv:0906.4123
\bibitem[\protect\citeauthoryear{DES Collaboration}{2005}]{2005astro.ph.10346T}
Dark Energy Survey Collaboration, 2005, arXiV:0510346
\bibitem[\protect\citeauthoryear{Cunha}{2009}]{2009PhRvD..79f3009C}
Cunha C., 2009, Physical Review D, 79, 63009
\bibitem[\protect\citeauthoryear{D'Agostini}{2005}]{DAgostini:2005we}
D'Agostini G., 2005, arXiV:0511182
\bibitem[\protect\citeauthoryear{Fisher}{1935}]{Fisher}
Fisher R.A., 1935, J. Roy. Stat. Soc., 98, 39
\bibitem[\protect\citeauthoryear{Gull}{1989}]{Gull:1989uy}
Gull S.F., 1989, in Skilling J. (ed.), in ``Maximum entropy and Bayesian methods", Kluwer publishing, 511, 518
\bibitem[\protect\citeauthoryear{Hogg {et~al.}}{2010}]{Hogg}
Hogg D.W., Bovy J., Lang D., 2010, arXiV:1008.4686
\bibitem[\protect\citeauthoryear{Kelly}{2011}]{Kelly}
Kelly B.C., 2011, in Feigelson E., Babu J. (eds.), ``Statistical Challenges in Modern Astronomy V", Penn State, arXiV:1112.1745
\bibitem[\protect\citeauthoryear{Kim \& Miquel}{2007}]{Kim}
Kim A.G., Miquel R., 2007, Astroparticle Physics, 28, 448
\bibitem[\protect\citeauthoryear{Kitching et al.}{2008}]{Kitching2008}
Kitching T.D., Heavens A. F., Verde L., Serra P., Melchiorri A., 2008, PRD, 77, 3008
\bibitem[\protect\citeauthoryear{Mandel et al.}{2011}]{Mandel}
Mandel K.S., Narayan G., Kirshner R.P., 2011, ApJ, 731, 120
\bibitem[\protect\citeauthoryear{March {et~al.}}{2011}]{2011MNRAS.418.2308M}
March M.C., Trotta R., Berkes P., Starkman G.D., Vaudrevange P.M., 2011, MNRAS, 418, 2308
\bibitem[\protect\citeauthoryear{Refregier et al.}{2011}]{ARefregier:2011ho}
Refregier A., Amara A., Kitching T. D., Rassat A., 2011, A\&A, 528, 33
\bibitem[\protect\citeauthoryear{Schlegel et al.}{2011}]{BigBoss}
Schlegel D. et al., 2011, arXiV:1106.1706
\bibitem[\protect\citeauthoryear{Sellentin et al.}{2014}]{Sellentin}
Sellentin E., Quartin M., Amendola L., 2014, arXiV:1401.6892
\bibitem[\protect\citeauthoryear{Taylor {et~al.}}{1997}]{1997astro.ph..7265T}
Taylor A., Heavens A., Ballinger B., Tegmark M., 1997,  in ``Proceedings of the Particle Physics and Early Universe Conference'' (PPEUC), University of Cambridge, arXiv:9707265
\bibitem[\protect\citeauthoryear{Tegmark, Taylor, \& Heavens}{1997}]{Tegmark:1997}
Tegmark M., Taylor A., Heavens A., 1997, ApJ, 480, 22
\bibitem[\protect\citeauthoryear{Vogeley \& Szalay}{1996}]{VS96}
Vogeley, M., Szalay A., 1996, ApJ, 465, 34
\bibitem[\protect\citeauthoryear{Wolz et al.}{2012}]{Wolz}
Wolz L. et al., 2012, JCAP, 9, 009
\bibitem[\protect\citeauthoryear{Woodbury}{1950}]{1950woodbury}
Woodbury M.A., 1950, Statistical Research Group, Memo Rep. No. 42
\end{thebibliography}


\end{document}